\documentclass[prb,twocolumn,showpacs,floatfix,amsfonts]{revtex4}
\usepackage{graphicx,graphics,color,epsfig}
\usepackage{bm}
\usepackage{amsmath}
\usepackage{amssymb}
\usepackage{epstopdf}
\usepackage{fancyheadings}
\begin{document}
\title{Vortex core states in a minimal two-band model for iron-based superconductors}
\author{Xiang Hu and C. S. Ting}
\affiliation{Texas Center for Superconductivity and Department of Physics, 
University of Houston,  Houston, Texas 77204}
\author{Jian-Xin Zhu}
\email[Corresponding author. \\ Electronic address: ]{jxzhu@lanl.gov}
\homepage{http://theory.lanl.gov}
\affiliation{Theoretical Division, Los Alamos National Laboratory,
Los Alamos, New Mexico 87545}
\date{\today}
\begin{abstract}
The pairing symmetry is one of the major issues in the study of
iron-based superconductors. We adopt a minimal two-band tight-binding model with various channels of 
pairing interaction, and derive a set of two-band Bogoliubov-de Gennes (BdG) equations.
The BdG equations are implemented in real space and then solved self-consistently via exact diagonalization. In the uniform case, we find that the $d_{x^2-y^2}$-wave pairing state is most favorable 
for a nearest-neighbor pairing interaction while the $s_{x^2y^2}$-wave pairing state is most favorable for a next-nearest-neighbor pairing interaction. The is consistent with that reported by Seo {\em et al.}  [Phys. Rev. Lett. {\bf 101}, 206404 (2008)].
We then proceed to study the local electronic structure around a magnetic vortex  core for both $d_{x^2-y^2}$-wave and $s_{x^2y^2}$-wave pairing symmetry in the mixed state. It is found from the local density of states (LDOS) spectra and its spatial variation that the resonance core states near the Fermi energy for the $d_{x^2-y^2}$-wave pairing symmetry are bound while those for the $s_{x^2y^2}$-wave pairing symmetry can evolve from the localized states into extended ones with varying electron filling factor. Furthermore, by including an effective exchange interaction, the emergent antiferromagnetic 
spin-density-wave (SDW) order can suppress the resonance core states, which provides one possible avenue to understand the absence of resonance peak as revealed by recent scanning tunneling microscopy experiment (STM) by Yin {\em et al.} [Phys. Rev. Lett. {\bf 102}, 097002 (2009)].  
\end{abstract}
\maketitle

\section{Introduction}
\label{SEC:Intro}

Recent discovery of the iron-based 
superconductors~\cite{YKamihara08,XHChen08,GFChen08a,ZARen08a,ZARen08b,MRotter08a,MRotter08b,CKrellner08,GFChen08b,QHuang08,GWu08,JZhao08,NNi08,GFChen08c,ZRen08,HSJeevan08,FCHsu08,KWYeh08,XCWang08,JHTapp08,MJPitcher08,DRParker09} has generated considerable interest in the condensed community.~\cite{MRNorman08} It is the only class of non-cuprate materials with high 
superconducting transition temperature.  The iron-based superconductors bears both similarity to and important differences from cuprate superconductors. On the one hand, both families of superconductors have a layered structure and exhibit magnetism in undoped phase. On the other hand, although the undoped cuprate superconductors are antiferromagnetic Mott insulators, almost all iron-based parent  compounds are still metals, which have an SDW instability with a small magnetic moment.~\cite{CdelaCruz08,MAMcGuire08} In addition, band structure calculations~\cite{DJSingh08,GXu08,KHaule08} based on the local density approximation to density functional theory have emphasized the multi-orbital nature in the new superconductors, 
in contrast to the cuprates, where only the $d_{x^2-y^2}$-orbital is most important.
Now it seems commonly accepted from these calculations that the Fermi surface consists mainly of two electron sheets and two hole sheets in most of the parent compounds. 
 
To uncover the mechanism of superconductivity in these materials, the determination of pairing symmetry of the superconducting order parameter is a good starting point. 
Up to now, many types of  pairing symmetry have been 
suggested, ranging from  the $p$-wave symmetry,~\cite{PALee08} 
mixed $d$-wave symmetry,~\cite{JLi08, QSi08} and extend $s$-wave 
symmetry,~\cite{KSeo08,MMParish08} besides the $\pm s$-wave symmetry.~\cite{IIMazin08}  
Experimentally,  the pairing symmetry can 
be either revealed by such phase-sensitive methods as multi-crystal junctions,~\cite{MSigrist95,DJvanHarlingen95,CCTsuei00}  or inferred by non-phase-sensitive techniques like angled-resolved photoemission spectroscopy~\cite{ADamascelli03} and tunneling,~\cite{OFischer07} and other more traditional thermodynamical measurements.  The first type of technique has not been applied successfully to the iron pnictides yet. The latter type of techniques are making strides by providing detailed information about the nature of single-particle excitations. At this moment, the consensus of the pairing symmetry in the new class of superconductors remains unsettled in view of the fact that conflicting data have been reported.  For example, results from some tunneling,~\cite{LShan08,YWang09} photoemission,~\cite{TSato08} and nuclear-spin-lattice relaxation~\cite{SKawasaki08,KMatano08,HJGrafe08,HMukuda08,YNakai08} measurements indicate a $d$-wave pairing symmetry while those from other tunneling,~\cite{TYChen08} photoemission,~\cite{HDing08,TKondo08,LWray08} specific heat,~\cite{GMu08} and penetration depth~\cite{LMalone08,KHashimoto09,CMartin09} measurements show the 
evidence for an $s$-wave pairing symmetry. 

To identify the pairing symmetry being conventional or unconventional, the electronic structure around local inhomogeneity such as single impurities/defects, surfaces/interfaces, and magnetic vortices can provide useful information,~\cite{AVBalatsky06,OFischer07,HAlloul09} which can be measured by such local probes as STM and nuclear magnetic resonance. Recently, the local electronic states near impurities~\cite{YBang08} and surfaces/interfaces~\cite{HYChoi08,PGhaemi09,JLinder09,WFTsai08,AAGolubov08,SOnari09,MANAraujo09} have been studied in the context of $s_{\pm}$-wave~\cite{IIMazin08} or $s_{x^2y^2}$-wave~\cite{KSeo08} pairing symmetry in a two-band model for the iron-based superconductors, where the existence of impurity resonance or surface Andreev bound states is discussed. In a single-band model as relevant to high-$T_c$ cuprates, 
earlier studies~\cite{YWang95,MFranz98} have revealed the difference of local electronic states in the mixed state between $s$-wave and $d$-wave pairing symmetry.  The purpose of this paper is to report a comparison study of local electronic structure in the mixed state for both the $d_{x^2-y^2}$-wave and $s_{x^2y^2}$-wave pairing symmetry within a minimal two-band model of superconductivity.

The outline of the paper is as follows. In Sec.~\ref{SEC:Method}, we introduce the multi-band model for the iron-pnictide superconductor and derive a set of multi-band BdG equations. In 
Sec.~\ref{SEC:Results}, we present numerical results first for the uniform case to identify the  favorable pairing symmetry for a given channel of pairing interaction, and then for the mixed state to investigate the local electronic structure for both $d_{x^2-y^2}$-wave and $s_{x^2y^2}$-wave pairing symmetry. 
The effect of the emergent antiferromagnetic SDW order on the resonance core states is also discussed. 
Finally, a summary is given in Sec.~\ref{SEC:Summary}.

\section{Bogoliubov-de Gennes Equations for Multi-Band Superconductors}
\label{SEC:Method}

We start with a multi-band model for the normal state 
band structure and introduce the phenomenological 
pairing interactions within each band. 
The Fermi surface topology as mapped out by the magneto-oscillation measurements consists 
of weakly corrugated small-size cylinders. Therefore, we consider only a two-dimensional (2D) 
system  for simplification. The magnetic field applied along the direction perpendicular to the 
2D plane creates quantized vortices in the mixed state.
The model Hamiltonian can be written as:
\begin{eqnarray}
\mathcal{H} &=& -\sum_{ij,\alpha\beta,\sigma} (\tilde{t}_{ij,\alpha\beta} + \mu \delta_{ij}
\delta_{\alpha\beta} ) c_{i\alpha\sigma}^{\dagger} c_{j\beta\sigma} \nonumber \\
&& + \sum_{ij,\alpha\beta}
[\Delta_{ij,\alpha\beta} c_{i\alpha\uparrow}^{\dagger} c_{j\beta\downarrow}^{\dagger} +\text{h.c.}]\;. 
\label{EQ:Hamil}
\end{eqnarray}
Here the operators $c_{i\alpha\sigma}$ ($c_{i\alpha\sigma}^{\dagger}$) annihilate (create) an electron at the $i$-th site in the orbital $\alpha$ and of the spin projection $\sigma$.  In the presence of the magnetic field, when the electron hops from the $j$-site to the $i$-site, a Peierls phase factor is acquired such that 
$
\tilde{t}_{ij,\alpha\beta} = t_{ij,\alpha\beta} \exp[i\frac{\pi}{\Phi_0} \int_{j}^{i} \mathbf{A}\cdot d\mathbf{r}]$,
where $t_{ij,\alpha\beta}$ is the zero-field hopping integral and we assume it to be real, $\mathbf{A}$ is the vector potential, and the quantity  $\Phi_{0}=hc/2e$ is the superconducting magnetic flux quantum.
In the Hamiltonian~(\ref{EQ:Hamil}), the quantity $\mu$ is the chemical potential, and $\Delta_{ij,\alpha\beta}$ is the superconducting pair potential, which is given by 
$\Delta_{ij,\alpha\beta} = \delta_{\alpha\beta} V_{ij,\alpha} \langle c_{i\alpha \uparrow} c_{j\alpha\downarrow}\rangle$. In our model, only the spin singlet and intra-orbital pairing is considered.  However, we note that since our model considers one iron atom per unit cell, it is consistent with the picture of two electron sheets at the $M$ point and two hole sheets at the $\Gamma$ point obtained 
from the LDA calculations, where two iron atoms per cell are considered.

By introducing the canonical transformation:
\begin{equation}
c_{i\alpha\sigma} = \sum_{n(E_{n}>0)} [u_{i\alpha}^{n} \gamma_{n} -\sigma {v_{i\alpha}^{n}}^{*} \gamma_{n}^{\dagger}] \;, 
\end{equation}
we arrive at a set of multi-band BdG equations:
\begin{equation}
\sum_{j} \left[
\begin{array}{cc} 
\hat{H}_{ij} & \hat{\Delta}_{ij} \\
\hat{\Delta}_{ij}^{\dagger} & - \hat{H}_{ij}^{*}  
\end{array}
\right] 
\left[ \begin{array}{c}
\hat{u}_{j}^{n} \\ \hat{v}_{j}^{n} 
\end{array}
\right] = E_{n} \left[ \begin{array}{c}
\hat{u}_{i}^{n} \\ \hat{v}_{i}^{n} \end{array}
\right] \;.
\end{equation}
Here $(\hat{u}_{i}^{n}\;\; \hat{v}_{i}^{n})^{Transpose}$ are the eigenstates corresponding eigenenergies $E_{n}$.  The variables with the symbol  ``\^{}''  mean matrices or vectors in the orbital space,
with the single particle Hamiltonian
\begin{equation}
H_{ij,\alpha\beta} = -\tilde{t}_{ij,\alpha\beta} -\mu \delta_{ij}\delta_{\alpha\beta} \;,
\end{equation}
and the pair potential subject to the self-consistency condition 
\begin{eqnarray}
\Delta_{ij,\alpha\beta} &=& \frac{\delta_{\alpha\beta}V_{ij,\alpha}}{2} \sum_{n(E_{n}>0)} (u_{i\alpha}^{n} {v_{j\alpha}^{n}}^{*} + u_{j\alpha}^{n} {v_{i\alpha}^{n}}^{*} ) \nonumber \\
&&\times \tanh\biggl{(} \frac{E_{n}}{2k_{B}T} 
\biggr{)}\;.
\end{eqnarray}
Notice that the qusiparticle excitation energies are measured with respect to the Fermi energy.

We solve the above set of BdG equations self-consistently: First guess an initial pair potential $\Delta_{ij}$ and exactly diagonalize the equation; use the obtained eigenfunctions and eigenvalues to update the pair potential; repeat the procedure until the desired convergence criterion is satisfied.
Once the self-consistency is achieved, the LDOS is calculated as:
\begin{eqnarray}
\rho_{i}(E) &=& \frac{2}{N_{c}} \sum_{\mathbf{K},\alpha,n(E_n>0)} 
\biggl{[}\vert u_{i,\alpha}^{n}(\mathbf{K})\vert^{2} \delta(E-E_{n}(\mathbf{K}))  \nonumber \\
&&+ \vert v_{i,\alpha}^{n}(\mathbf{K})\vert^{2} \delta(E+E_{n}(\mathbf{K}))\biggr{]} \;,
\label{EQ:LDOS}
\end{eqnarray}
where the factor 2 accounts for the spin degeneracy and $N_c$ is the number of magnetic unit cells.
This quantity is proportional to the differential tunneling conductance as measured by STM.~\cite{MTinkham96}

\begin{figure}
\includegraphics[width=8cm]{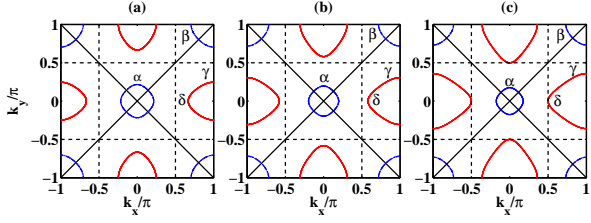}
\caption{(Color online) Fermi surface in the Brillouin zone with one Fe per cell for various values of chemical potential $\mu=1.6$ (a), $1.8$ (b), and $2.0$ (c). The diagonal solid lines are the nodal lines for $d_{x^2-y^2}$-wave pairing symmetry while the horizontal and vertical dashed lines are the nodal lines for $s_{x^2y^2}$-wave pairing symmetry. The sheets at the zone center and corners are hole pockets, and those at the zone horizontal and vertical boundaries are electron pockets. }
\label{FIG:FS}
\end{figure}

The established set of the BdG equations is general  for describing quasiparticle excitations of a superconductor with any number of bands. It is applicable to any version of the multi-band tight-binding  model as recently proposed for the low-energy $d$-electron physics for the iron-based superconductors.~\cite{SRaghu08,TLi08,QHan08,YRan09,PALee08,CCao08,KKuroki08} 
Several groups~\cite{PALee08,CCao08,KKuroki08} have pointed out that one may need at least three orbitals to accurately reproduce the LDA band structure and Fermi surface topology.  
However, it has also been argued recently~\cite{SRaghu08,YRan09} 
that the Fe 3$d_{xz}$ and 3$d_{yz}$orbitals play an important role in the low energy physics of these materials.  
For simplicity, we consider in the present paper a minimal two-band model as suggested in Ref.~\onlinecite{SRaghu08}, with the following tight-binding hopping integrals 
$t_1=-1.0$, $t_2=1.3$, $t_3=t_4=-0.85$. These hopping parameters
appear in the normal-state energy dispersion in the unfolded Brillouin zone as
\begin{equation}
E_{\mathbf{k}} = \frac{\epsilon_{11}+\epsilon_{22}}{2} \pm 
\sqrt{\biggl{(} \frac{\epsilon_{11} 
-\epsilon_{22}}{2}\biggr{)}^{2} + \epsilon_{12} \epsilon_{21}}\;,
\end{equation}
where 
\begin{eqnarray*}
\epsilon_{11}(\mathbf{k}) &=& -2t_{1} \cos k_x - 2t_2 \cos k_y -4 t_3 \cos k_x \cos k_y -\mu\;, \\
\epsilon_{22}(\mathbf{k}) &=& -2t_{2} \cos k_x - 2t_1 \cos k_y -4 t_3 \cos k_x \cos k_y-\mu\;, \\
\epsilon_{12}(\mathbf{k}) &=& \epsilon_{21}(\mathbf{k}) = -4t_{4} \sin k_x \sin k_y \;.
\end{eqnarray*}
The chemical potential $\mu=1.54$ corresponds to the half-filled case. In the rigid band approximation, we will relax this parameter $\mu$ to model the superconducting state upon the electron doping. 
The Fermi surface for various values of chemical potential is shown in Fig.~\ref{FIG:FS}. As expected, the increase of chemical potential, which is equivalent to the increase of the electron filling factor, enlarges the electron pockets but shrinks the hole pockets.
Throughout the paper, the energies are measured in units of  $|t_1|$. The temperature is set to be $T=0.01$ for the self-consistency calculations. We further choose the identical pairing interaction for each band and denote $V_{nn}$ as the nearest-neighbor pairing interaction strength and $V_{nnn}$ as the next-nearest-neighbor pairing interaction strength. 
For the calculation of the density of states,  the Dirac delta-function appearing in Eq.~(\ref{EQ:LDOS}) is approximated by 
\begin{equation}
\delta(E-E^{\prime}) \rightarrow \frac{1}{\pi} \frac{\Gamma}{(E-E^{\prime})^{2} + \Gamma^{2}}\;,
\end{equation}
with $\Gamma=0.01$ being chosen.

\section{Numerical Results and Discussions}
\label{SEC:Results}


We first sort out the most favorable pairing symmetry in the absence of magnetic field. 
For the nearest-neighbor pairing interaction $V_{nn}$ only,  the $d_{x^2-y^2}$-wave ($\propto \cos k_x - \cos k_y$)  pairing symmetry 
is more favorable
than the $s_{x^2+y^2}$-wave ($\propto \cos k_x + \cos k_y$) pairing symmetry. In detail, no matter whether the initial input
order parameter is of $d_{x^2-y^2}$-wave type or $s_{x^2+y^2}$-wave type, the converged solution 
will be dominantly of $d_{x^2-y^2}$-wave type. Similarly, for the next-nearest-neighbor pairing interaction $V_{nnn}$ only, 
the $s_{x^2 y^2}$-wave ($\propto \cos k_x  \cos k_y$) pairing  symmetry  is more favorable over the $d_{xy}$-wave ($\propto \sin k_x  \sin k_y$) type.
This observation is also consistent with earlier work.~\cite{KSeo08}

\begin{figure}
\includegraphics[width=1\linewidth]{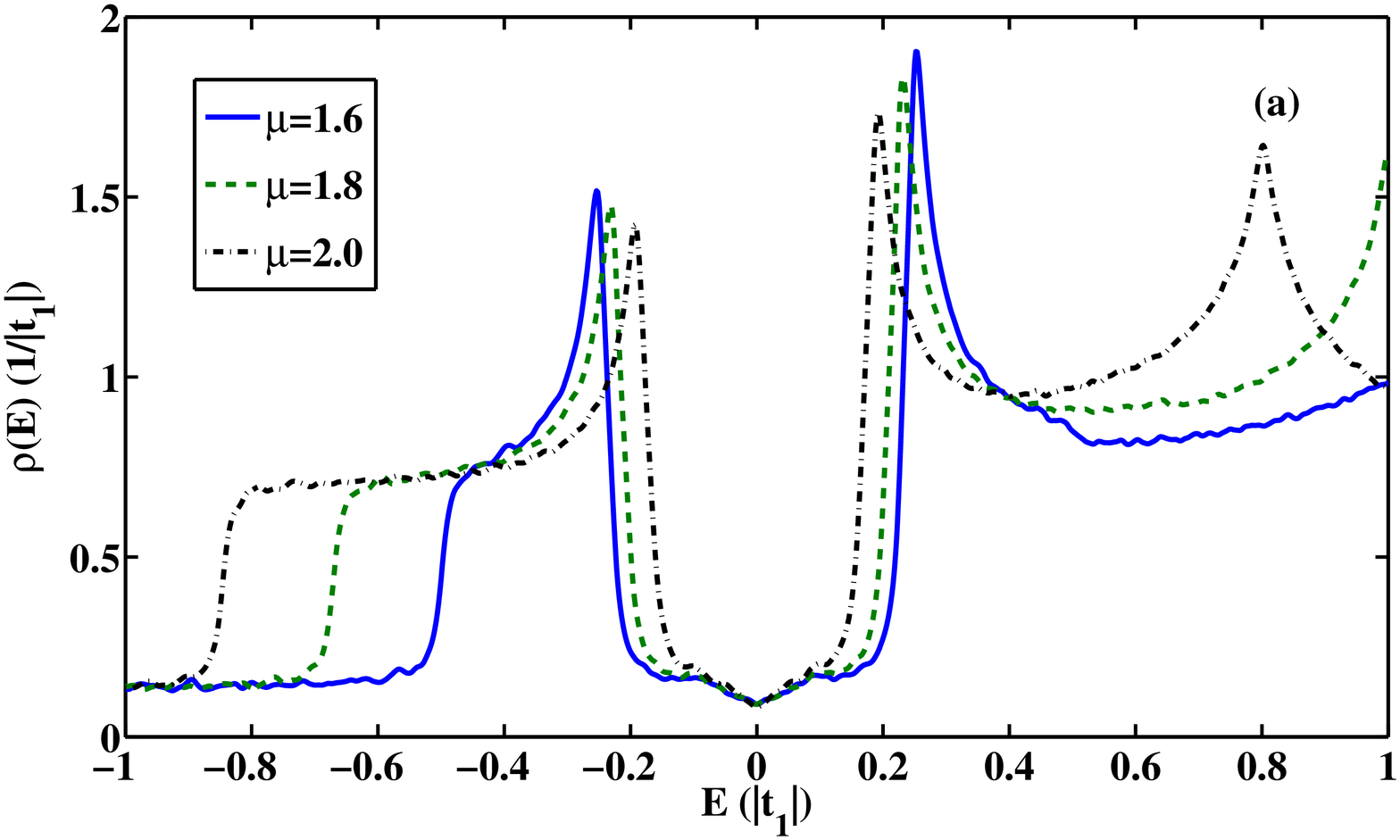}
\includegraphics[width=1\linewidth]{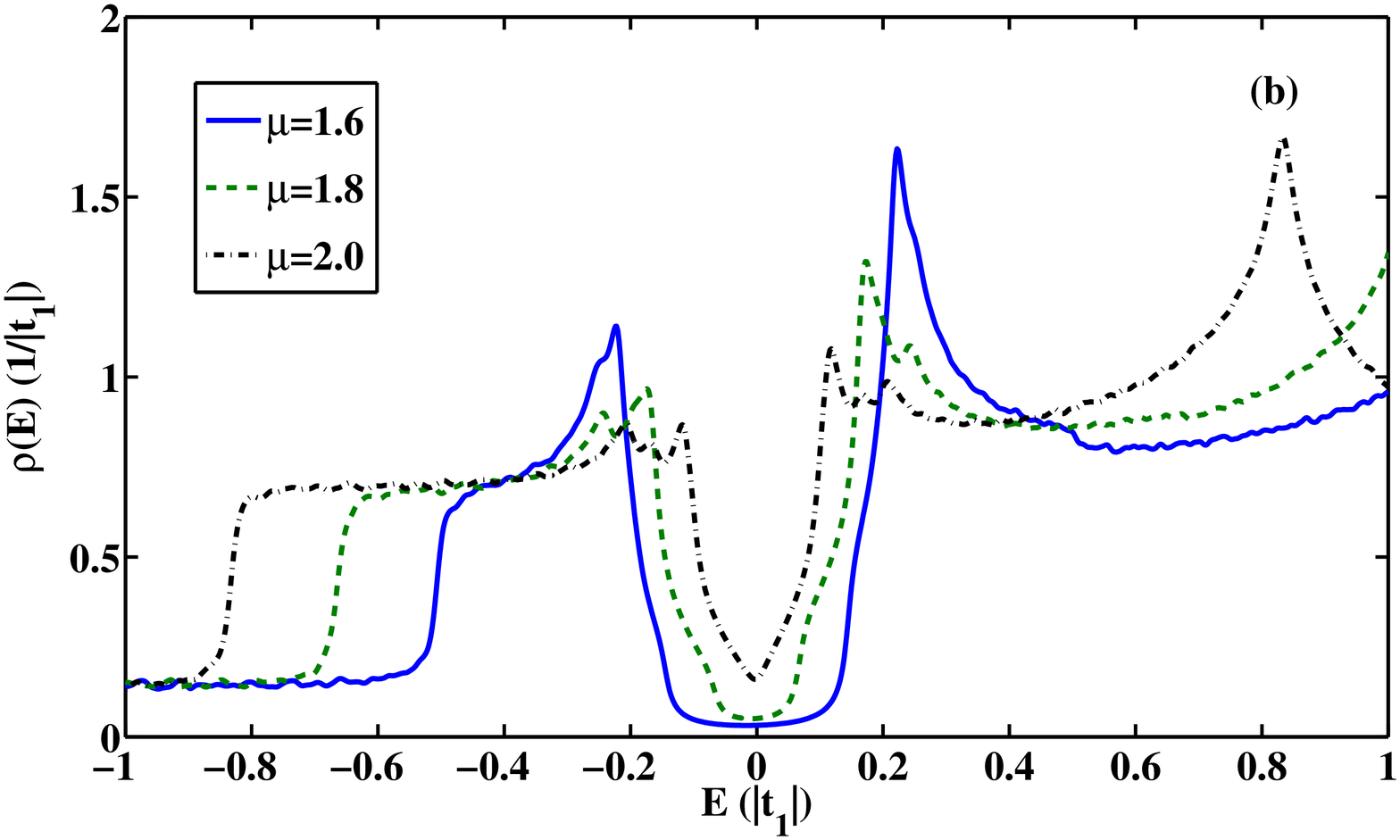}
\caption{(Color online) The bulk density of states as a function of energy in a uniform form system with a $d_{x^2-y^2}$-wave pairing symmetry (a) and an $s_{x^2 y^2}$-wave pairing symmetry for various values of chemical potential 
$\mu$. The pairing interaction for the corresponding channel is chosen to be $V_{nn}=1.5$ or $V_{nnn}=1.5$. }
\label{FIG:Bulk-DOS}
\end{figure}

The typical density of states (DOS) for these types of pairing symmetry is shown in Fig.~\ref{FIG:Bulk-DOS}.
For both types of pairing symmetry, the DOS shows a well-defined coherent peaks with the position determined by the order parameter. The additional peak outside the coherent peaks is due to the van Hove singularity, which is a normal state property because its location is fixed in the band structure.
It shows that the DOS feature in the multi-band model can be very different than that in the single band model. Specifically, for the $d_{x^2-y^2}$-wave pairing symmetry, the resultant DOS exhibits a linear behavior within a very small energy range near the Fermi energy and then reaches a plateau up to a larger energy scale characterized by the amplitude of the $d$-wave order parameter. For $s_{x^2y^2}$-wave pairing symmetry, the DOS feature is sensitive to the doping, and can have a $V$-shape when 
the electron pockets are enlarged to cross into the region in the Brillouin zone where the $s_{x^2y^2}$-wave order parameter has a sign change (see Fig.~\ref{FIG:Bulk-DOS}(b) and Fig.~\ref{FIG:FS}(c) for $\mu=2.0$). 

In the following, we consider the local electronic structure around a vortex core for both the $d_{x^2-y^2}$-wave and $s_{x^2y^2}$-wave pairing symmetry in the mixed state. 
For this purpose, the amplitude of the 
magnetic field is determined by the condition that each magnetic unit cell contains two superconducting flux quanta, i.e., $H=2\Phi_0/(N_x N_y a^{2})$, where $N_x$ and $N_y$ are the linear dimension of a square lattice with lattice constant $a$. We choose $N_x = 2N_y$ with $N_y$ being typically 16 and 20.
The number of magnetic unit cells is chosen to be $N_c = M_x M_y$ with $M_x=N_y$ and 
$M_y = N_x$, or $M_x=N_y/2$ and 
$M_y = N_x/2$. 

\begin{figure}
\includegraphics[width=1\linewidth]{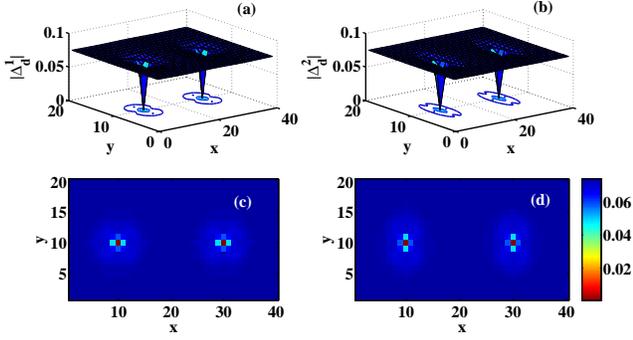}
\caption{(Color online)
The surface and imaging plot of the variation of $d_{x^2-y^2}$-wave order 
parameter around the magnetic vortices for band 1 ((a) and (c)) and band 2 ((b) and (d)).
Here the chemical potential $\mu = 1.60$, 
which is corresponding to the slightly electron doping case.}
\label{FIG:D-OP}
\end{figure}

\begin{figure}
\includegraphics[width=1\linewidth]{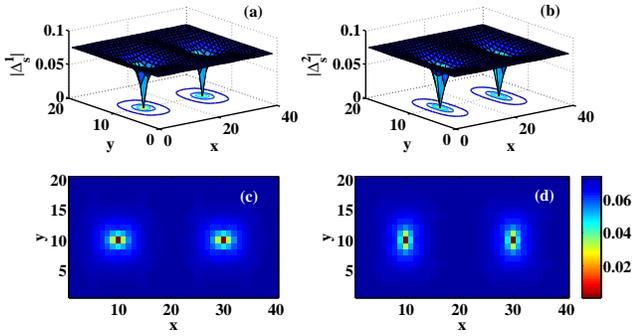}
\caption{(Color online) The surface and imaging plot of the variation of $s_{x^2 y^2}$-wave order 
parameter around the magnetic vortices for band 1 ((a) and (c)) and band 2 ((b) and (d)).
Here the chemical potential $\mu = 1.60$ is chosen.
}
\label{FIG:S-OP}
\end{figure}

\begin{figure}
\includegraphics[width=1\linewidth]{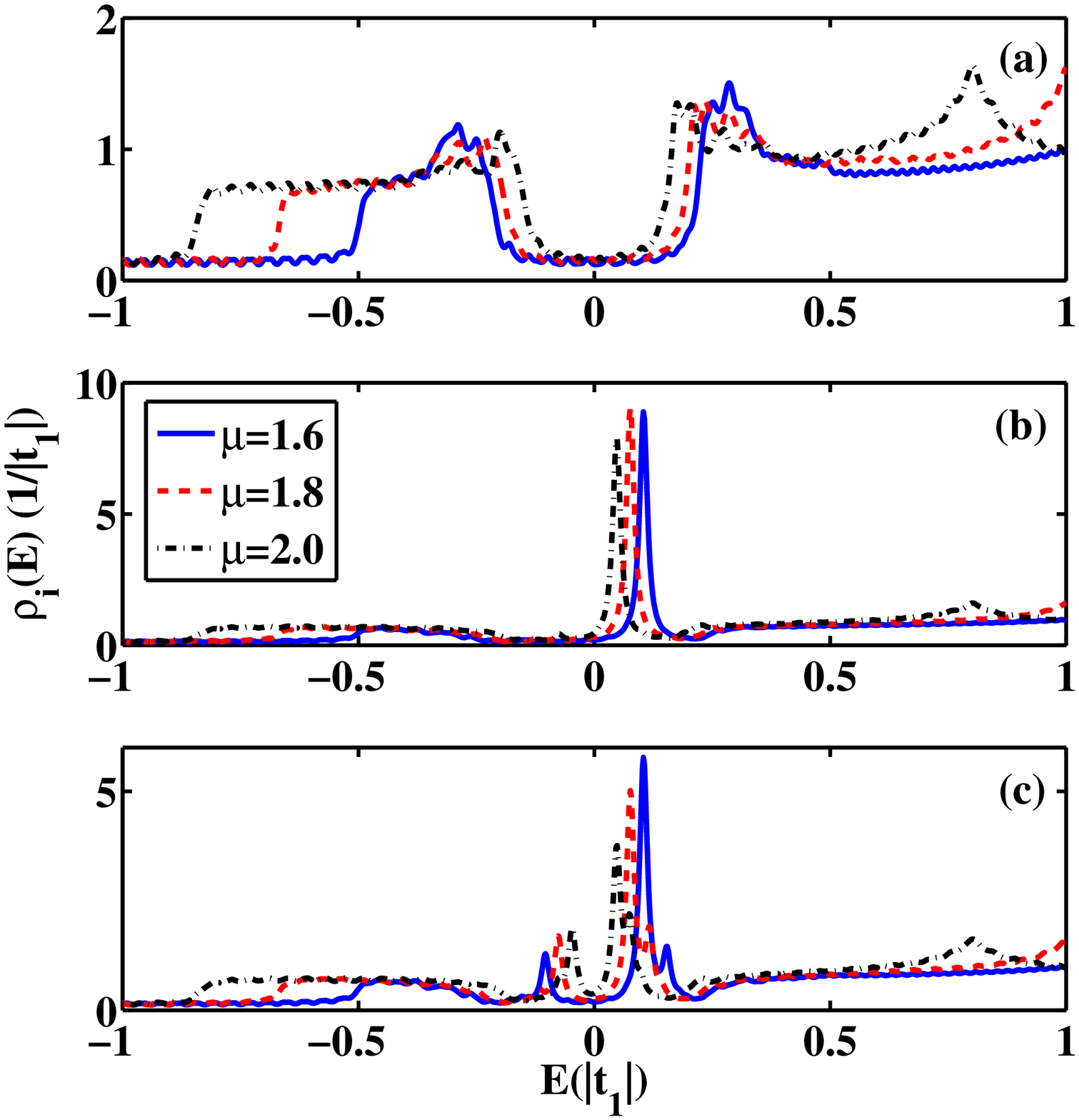}
\caption{(Color online) The local density of states as a function of energy for the  $d_{x^2-y^2}$-wave  model at at the center of the magnetic unit cell (a), at the
vortex core center (b), and at one of its nearest neighboring sites (c)  for various values of chemical potential.}
\label{FIG:D-LDOS}
\end{figure}

\begin{figure}
\includegraphics[width=1\linewidth]{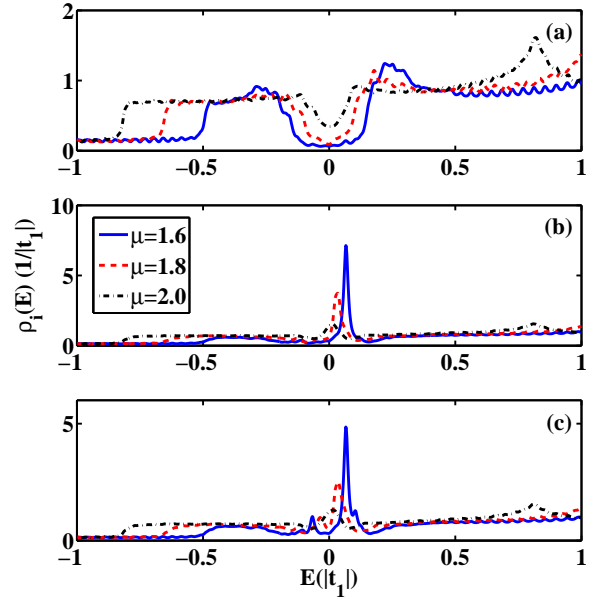}
\caption{(Color online) The local density of states as a function of energy for the $s_{x^2 y^2}$-wave 
model at the center of the magnetic unit cell (a),  at the vortex core center (b), and at  one of its nearest neighboring sites (c)  for various values of chemical potential.}
\label{FIG:S-LDOS}
\end{figure}

Representative self-consistent results for the order parameter of $d_{x^2-y^2}$-wave  and $s_{x^2y^2}$-wave pairing symmetry are shown in Figs.~\ref{FIG:D-OP} and \ref{FIG:S-OP}. The superconducting order parameter vanishes at the vortex center and starts to increase at the scale of superconducting coherence length to its bulk value. Since the hopping integrals along the $x$- and $y$-directions are nonequivalent within each band, the order parameter component associated with each individual band shows a two-fold symmetry. This is different from the case for the single-band model for high-$T_c$ cuprates, where the spatial dependence of $d$-wave order parameter shows a four-fold symmetry.   However, the profile of the order parameter associated with each individual band is related to that of the other band 
by a 90$^{\circ}$ rotation.

\begin{figure}
\includegraphics[width=1\linewidth]{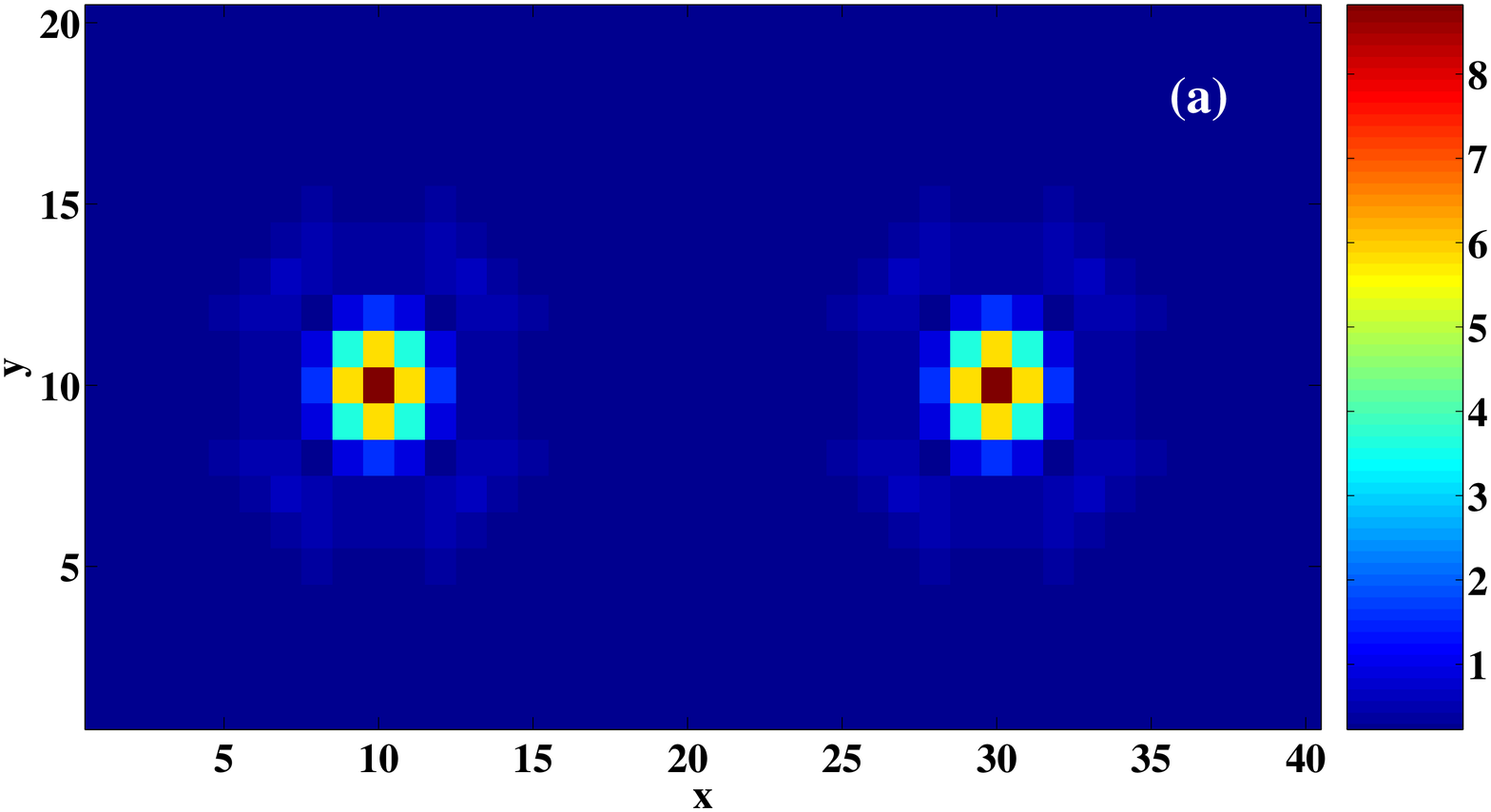}
\includegraphics[width=1\linewidth]{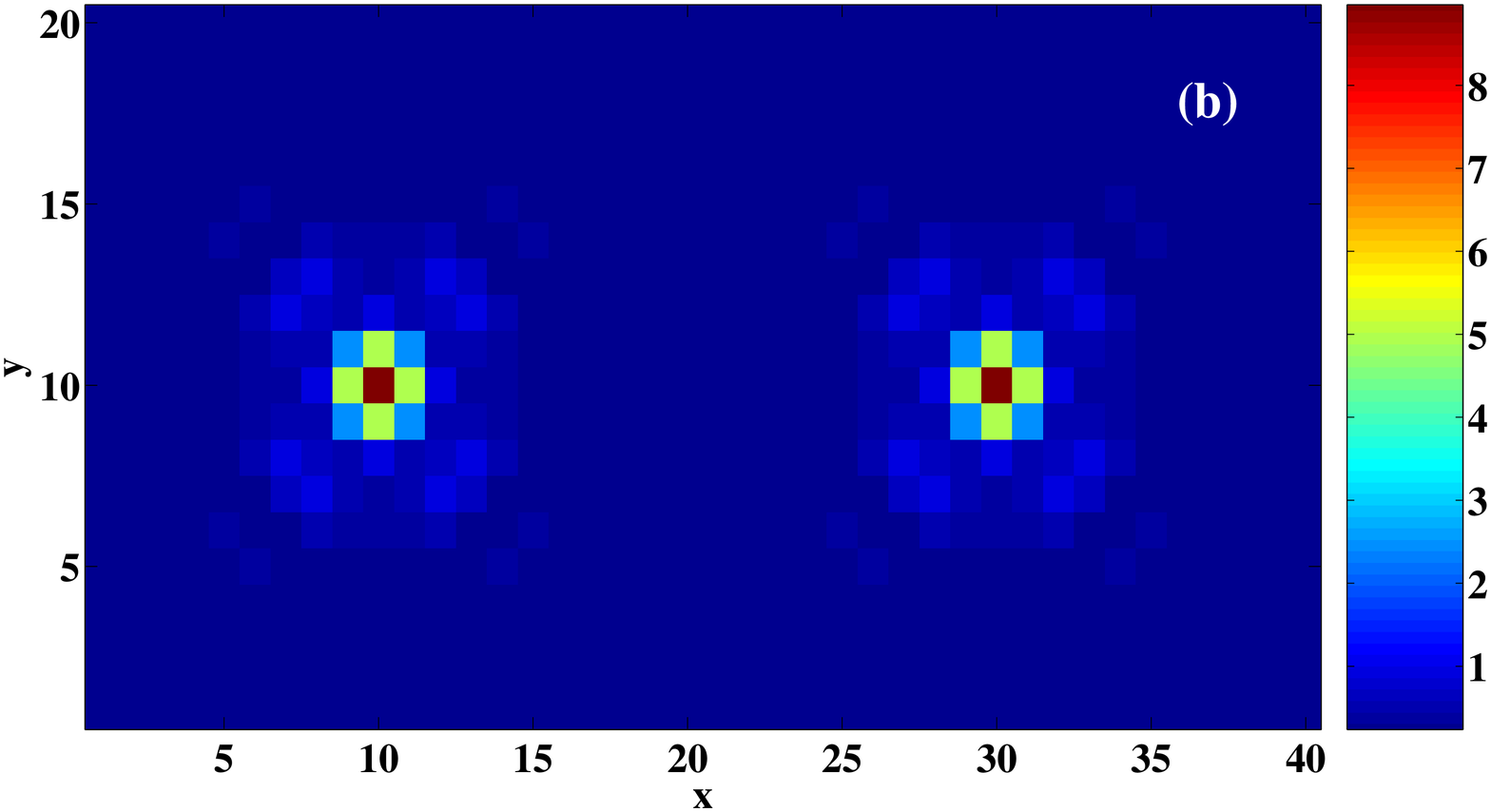}
\includegraphics[width=1\linewidth]{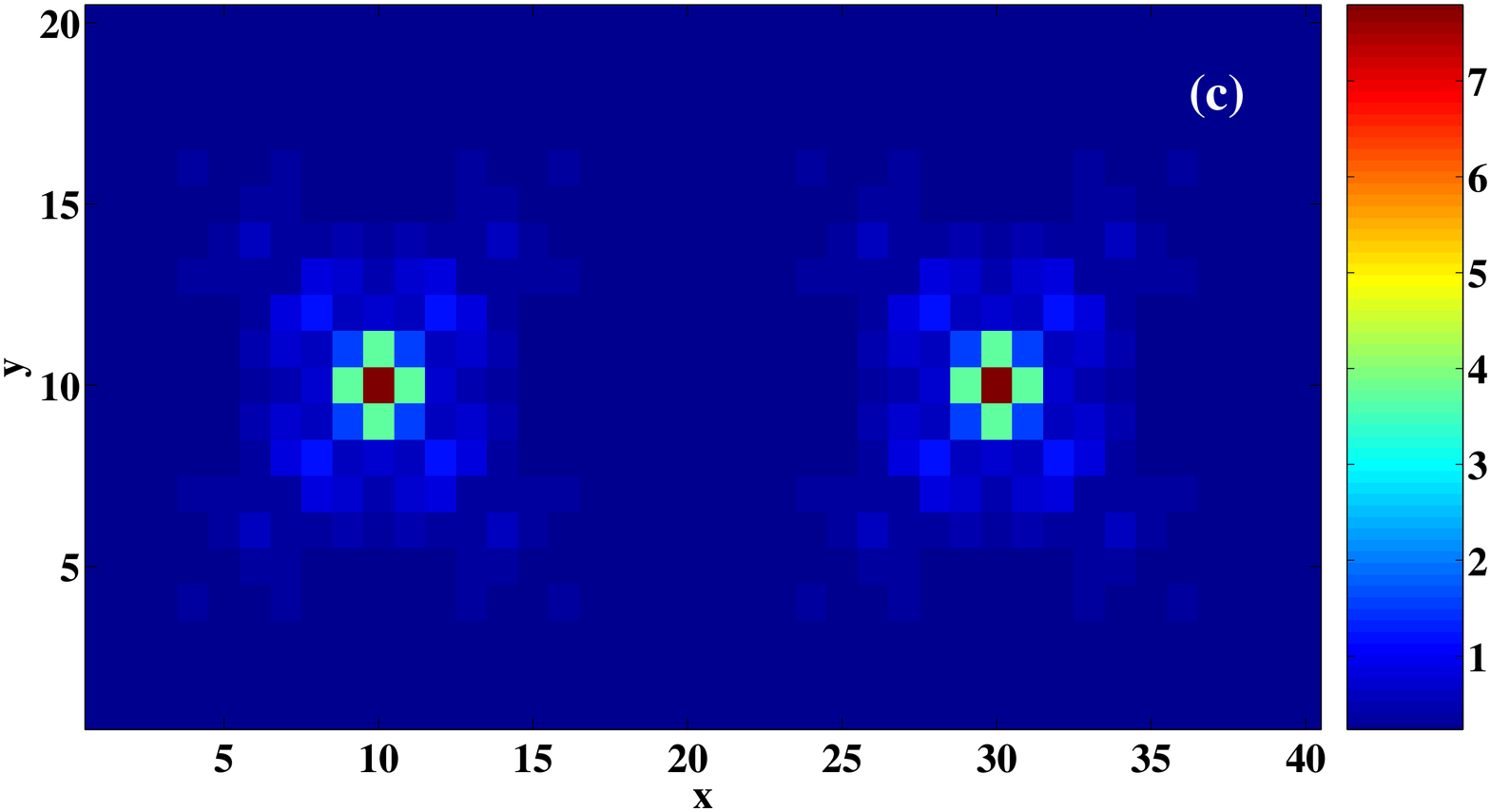}
\caption{(Color online) The imaging of the vortex core states at the resonance energy of the peak above the Fermi energy shown in Fig.~\ref{FIG:D-LDOS}(b) for the chemical potential values $\mu=1.6$ (a), 1.8 (b) and 2.0 (c).
\label{FIG:D-IMAG}}
\end{figure}

\begin{figure}
\includegraphics[width=1\linewidth]{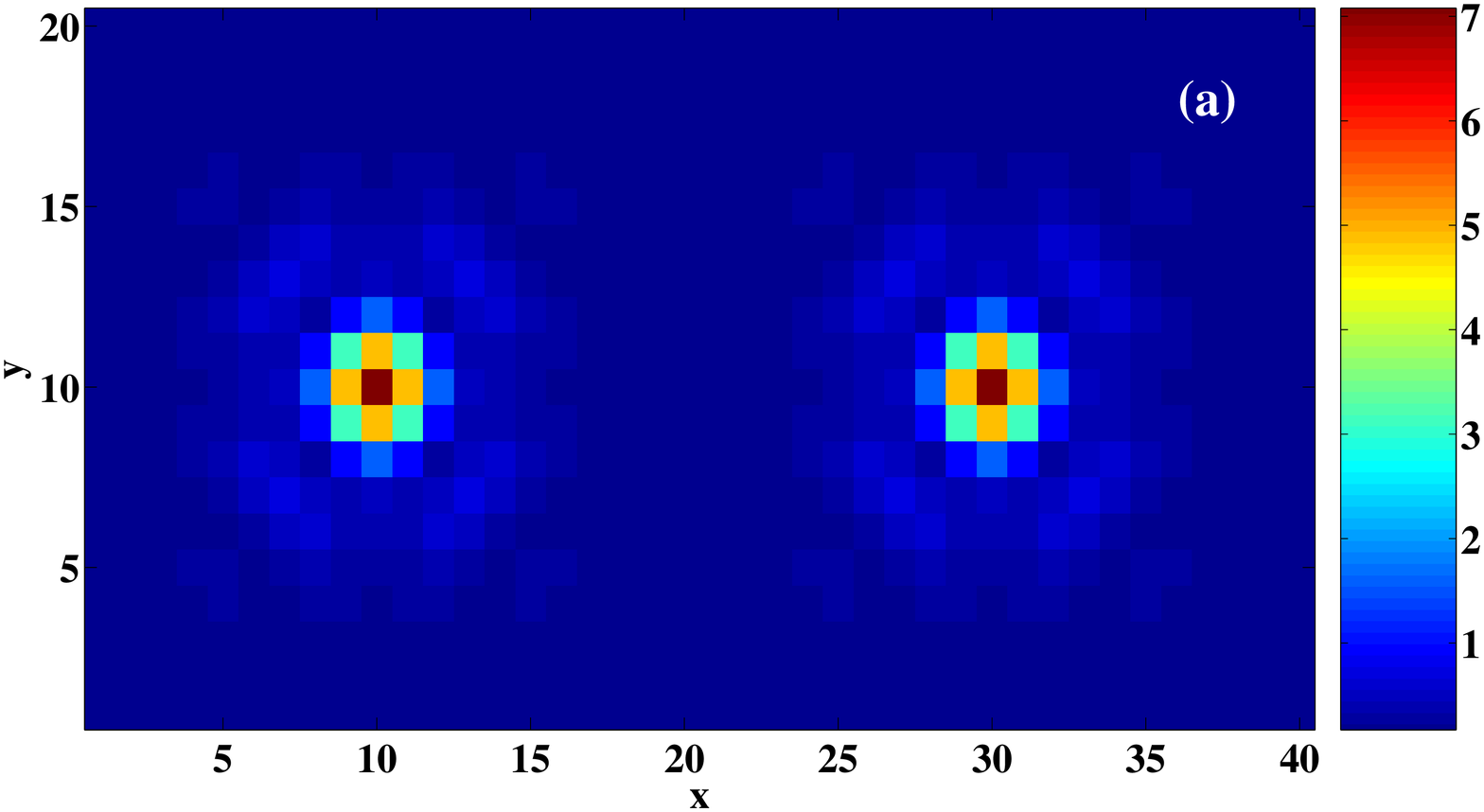}
\includegraphics[width=1\linewidth]{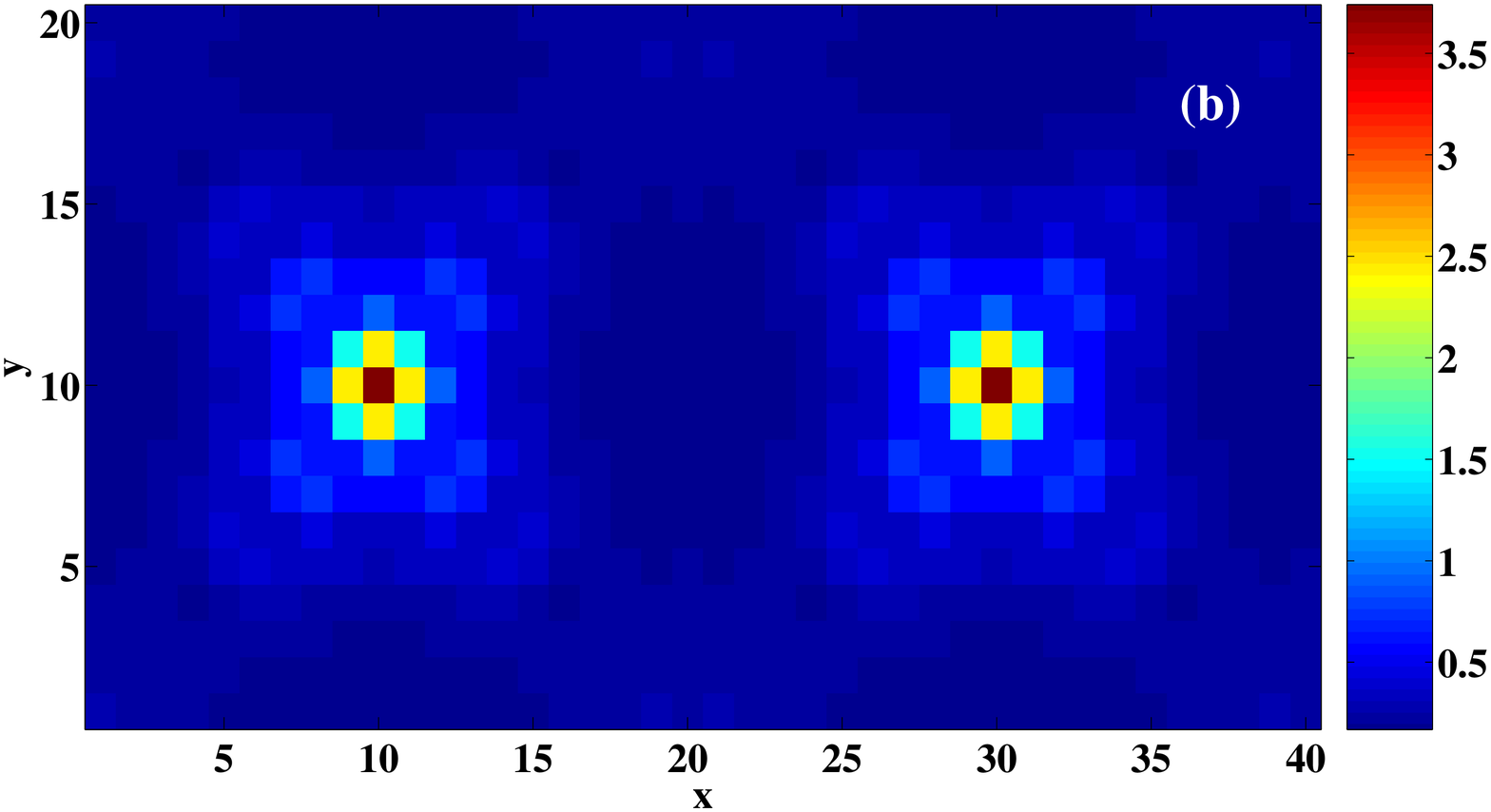}
\includegraphics[width=1\linewidth]{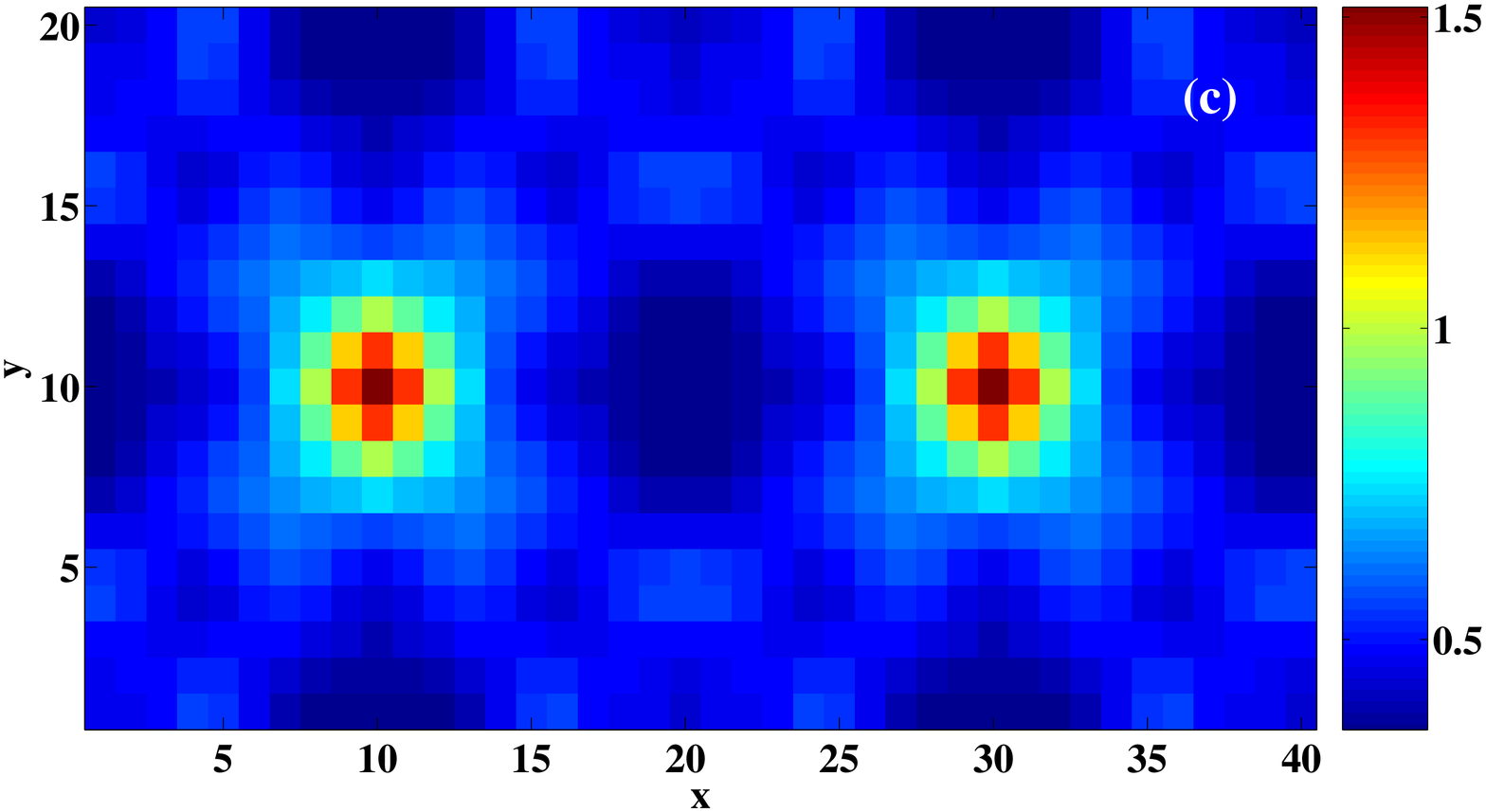}
\caption{(Color online) The imaging of the vortex core states at the resonance energy of the peak above the Fermi energy shown in Fig.~\ref{FIG:S-LDOS}(b) for the chemical potential values $\mu=1.6$ (a), 1.8 (b), and 2.0 (c).
\label{FIG:S-IMAG}}
\end{figure}

In Figs.~\ref{FIG:D-LDOS} and \ref{FIG:S-LDOS}, we show the LDOS  at a site midway between two nearest-neighboring vortices (i.e., the center of the magnetic unit cell), at the vortex core center, and at one of its nearest neighboring sites for both $d_{x^2-y^2}$-wave and $s_{x^2y^2}$-wave pairing symmetry. 
The LDOS at the midpoint between two nearest-neighboring vortices  exhibits a grossly similar feature to the case in the absence of magnetic field (Compare Figs.~\ref{FIG:D-LDOS}(a) with Fig.~\ref{FIG:Bulk-DOS}(a), and Fig.~\ref{FIG:S-LDOS}(a) with 
Fig.~\ref{FIG:Bulk-DOS}(b)). However, fine oscillating structure is observed in the flat region of density of states. By varying the size of the magnetic unit cell, we have numerically verified that the oscillation period is inversely proportional to the magnetic field, which suggests the fine structure is related to the Landau oscillation.
For both the $d_{x^2-y^2}$-wave and $s_{x^2y^2}$-wave pairing symmetry, two resonant peaks show  up near the Fermi energy in the LDOS at the vortex core center (see Figs.~\ref{FIG:D-LDOS}(b) and \ref{FIG:S-LDOS}(b)), which reflects the existence of Andreev bound states in the limit of isolated vortices. We note that, at the core center, the intensity of the peak located at the positive energy is much larger than that of the peak located at the negative energy. 
When the LDOS is measured near the core center (see Figs.~\ref{FIG:D-LDOS}(c) and \ref{FIG:S-LDOS}(c)), the resonance peak below the Fermi energy can be seen  more clearly.  The peaks are shifted toward the Fermi energy as the chemical potential is increased. Notably, the peaks in the LDOS for the $d_{x^2-y^2}$-wave pairing symmetry is not rigorously located at the Fermi energy, and the peak intensity only degrades sightly as the  chemical potential varies (see Fig.~\ref{FIG:D-LDOS}(b,c)). The peaks in the LDOS for the $s_{x^2y^2}$-wave pairing symmetry are broadened significantly as they are shifted toward the Fermi energy as the chemical potential is increased (see Fig.~\ref{FIG:S-LDOS}(b,c)).  These results suggest that, in the multi-band model, the vortex core states for the $d_{x^2-y^2}$-wave pairing symmetry are really bound states, while the vortex core states for the $s_{x^2y^2}$-wave pairing symmetry can be either localized or extended, depending on the chemical potential (i.e., the electron filling factor). The evidence for this observation is further enforced  by investigating the spatial variation of the LDOS at the positive resonance energy, as shown in Figs.~\ref{FIG:D-IMAG} and
\ref{FIG:S-IMAG}, where the resonance state for the $d_{x^2-y^2}$-wave pairing symmetry is bound around the core center while that for the $s_{x^2y^2}$-wave pairing symmetry begins to show long tails as the chemical potential is increased. The results are different from the single band model,~\cite{YWang95} where the resonance peak is much broadened in the LDOS for the $d_{x^2-y^2}$-wave pairing symmetry while it is much sharp for the $s$-wave pairing symmetry, demonstrating the sensitivity of quasiparticle properties to the electronic band structure.


%


\begin{figure}
\includegraphics[width=1\linewidth]{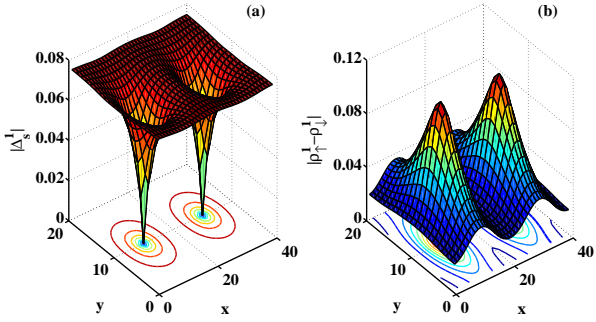}
\caption{(Color online) The spatial variation of the superconducting order parameter (a) and the absolute value of the SDW order parameter associated with an individual band for $J_{nn}=1.95$. The other parameter values are fixed at $V_{nnn}=1.5$ and $\mu=1.6$.}
\label{FIG:SDW-Profile}
\end{figure}

\begin{figure}
\includegraphics[width=1\linewidth]{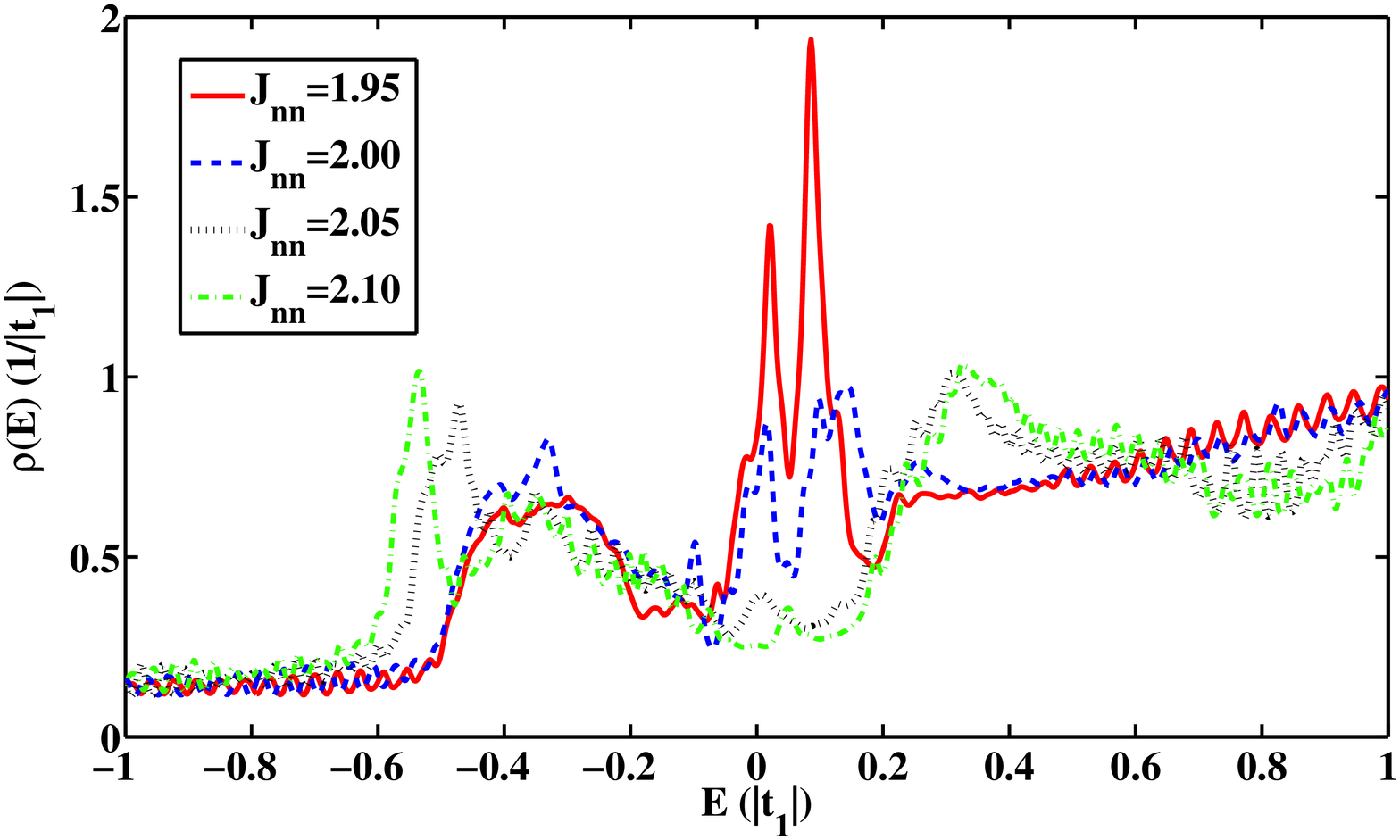}
\caption{(Color online) The local density of states as a function of energy at the $s_{x^2y^2}$-wave vortex core center for various values of exchange coupling strength $J_{nn}$. The other parameter values are fixed at $V_{nnn}=1.5$ and $\mu=1.6$.}
\label{FIG:SDW-LDOS}
\end{figure}

Experimentally, recent STM measurements~\cite{YYin09} 
on BaFe$_{1-x}$Co$_{x}$As$_{2}$ observed no zero-bias conductance 
peak at the vortex core center. On the one hand, this experimental finding is in striking contract to the theoretical results above for either  $d_{x^2-y^2}$-wave and $s_{x^2y^2}$-wave pairing symmetry. On the other hand, we notice that the superconductivity is in close proximity to the magnetism in iron pnictides. In particular, the temperature-doping electronic phase diagram shows that the antiferromagnetic SDW state is either coexistent with,~\cite{HChen09,AJDrew09} or incipiently interrupted by,~\cite{HLuetkens09}   or disappears immediately before~\cite{JZhao08b} the superconducting state. In the following, we introduce an Heisenberg-like exchange interaction term between the electronic spin density to take into account the magnetism. We will also restrict the consideration to the case of $s_{x^2y^2}$-wave pairing symmetry and assume the SDW order comes from the nearest neighbor spin-spin interaction. In the mean field approximation, this part of Hamiltonian can be written as:
\begin{equation}
H_{SDW}=  \sum_{i,\alpha} \Delta_{SDW,i,\alpha} (n_{i\alpha\uparrow}-n_{i\alpha\downarrow}) \;,
\end{equation}
where the SDW order parameter is given by 
\begin{equation}
\Delta_{SDW,i,\alpha} = \frac{1}{4}\sum_{\delta} J_{i,i+\delta}^{(\alpha)} m_{AF,i+\delta,\alpha}\;,
\end{equation}
with the magnetization $m_{AF,i,\alpha} = \langle n_{i\alpha\uparrow} \rangle - \langle 
n_{i\alpha\downarrow} \rangle$.
To enforce a collinear $(\pi,0)$ SDW state, we take the exchange coupling along the $x$- and $y$-direction to have opposite signs, which is consistent with the recent band structure 
calculations on LaOFeAs.~\cite{ZPYin08} We assume $J_{x}^{(\alpha)}=-J_{y}^{(\alpha)}=J_{nn}^{(\alpha)}$ with $J_{nn}^{(\alpha)}$ to be positive.  Although this assumption is oversimplified, it should still serve well to demonstrate  the effect of magnetic ordering. By including this term in the BdG equations, we repeat the numerical calculations for varying values of $J_{nn}$ ($J_{nn}^{(2)}=J_{nn}^{(1)}=J_{nn}$).   In Fig.~\ref{FIG:SDW-Profile}, we show a representative  spatial variation of superconducting order parameter and SDW order parameter associated with one of the bands for $J_{nn}=1.95$,  $V_{nnn}=1.5$, and $\mu=1.6$.  The SDW order parameter has maximum at the vortex core center and then decreases away from the core center, which means the SDW order is induced by the magnetic field and is nucleated at the vortex core center. In this case,  the value of the superconducting order parameter far away from the core center is almost unchanged as compared to the case in the absence of the magnetism. When the exchange coupling is increased, the SDW order becomes more spread while the superconducting order parameter becomes more suppressed accompanied by an expansion of the vortex core in size.   We note that for larger values of $J_{nn}$, the SDW can coexist with the superconducting order even in the absence of the magnetic field. The magnetic field  further suppresses the superconducting order parameter in the formation of magnetic vortices.  In Fig.~\ref{FIG:SDW-LDOS}, we show the LDOS as a function of energy for various values of the exchange coupling strength for the $s_{x^2y^2}$-wave pairing symmetry with fixed values of $V_{nnn}=1.5$ and $\mu=1.6$. We find that as the SDW order begins to nucleate at the vortex core, the resonance peak near the Fermi energy in the LDOS is split into a double-peak structure, and the overall intensity of the structure is decreased (see the curve for $J_{nn}=1.95$). With the increasing exchange coupling strength, the SDW order is further enhanced, and the peak is further split with decreased intensity (see the curve for $J_{nn}=2.00$). As the exchange coupling strength is sufficiently large, 
the resonance peak can be almost completely suppressed (see the curves for $J_{nn}=2.05$ and $2.10$). These results provide one possible account  for the absence of resonance peak at the vortex core center as revealed by the STM on the iron-pnictide superconductor BaFe$_{1-x}$Co$_{x}$As$_{2}$.

\section{Summary}
\label{SEC:Summary}
In summary, we have adopted a minimal two-band tight-binding model with various channels of 
pairing interaction, and derived a set of two-band BdG equations.
The BdG equations have been implemented in real space and then solved self-consistently via exact diagonalization. In the uniform case, it has been found that the $d_{x^2-y^2}$-wave pairing state is most favorable  for a nearest-neighbor pairing interaction while the $s_{x^2y^2}$-wave pairing state is most favorable for a next-nearest-neighbor pairing interaction. 
We have also studied the local electronic structure around a magnetic vortex  core for both $d_{x^2-y^2}$-wave and $s_{x^2y^2}$-wave pairing symmetry in the mixed state. It has been shown from the LDOS spectra and its spatial variation that the resonance core states near the Fermi energy for the $d_{x^2-y^2}$-wave pairing symmetry are localized while those for the $s_{x^2y^2}$-wave pairing symmetry can evolve from the localized states into extended ones with varying electron filling factor. Furthermore, by including an effective exchange interaction, we have shown that the emergent antiferromagnetic SDW order can suppress the resonance core states. The emergence of the antiferromagnetic SDW states, regardless of being field induced or pre-existent  provides one possible avenue to understand the absence of resonance peak as revealed by recent STM experiment.  Further studies are still necessary to finally pin down the origin of  the missing resonance vortex core state in iron-based superconductors. 

\acknowledgments
We thank A. V. Balatsky, J. C. Davis, M. Graf, J. E. Hoffman, Jiangping Hu, V. Madhavan, Shuheng Pan, and Qimiao Si for helpful discussions.  We also thank Degang Zhang and T. Zhou for discussion and collaboration on related project. One of the authors (X.H.) acknowledges the hospitality of Los Alamos National Laboratory (LANL), where this work was initiated. We acknowledge the U.S. DOE CINT at LANL for computational support.  This work was supported by the Robert Welch Foundation No. E-1146 at the University of Houston (X.H. and C.S.T.), by U.S. DOE at LANL under Contract No. DE-AC52-06NA25396, the U.S. DOE Office of Science, and the LANL LDRD Program (J.X.Z.).

\end{document}